\documentclass[11pt]{article}

\usepackage[a4paper,margin=1in]{geometry}
\usepackage{fontspec}        
\usepackage{xeCJK}           
\usepackage{microtype}

\usepackage{amsmath,amssymb,amsthm,mathtools}

\usepackage{graphicx}
\graphicspath{{figures/}}
\usepackage{booktabs}
\usepackage{array}
\usepackage{multirow}
\usepackage{subcaption}
\usepackage{algorithm}
\usepackage{algpseudocode}

\usepackage[numbers,sort&compress]{natbib}
\usepackage{hyperref}
\usepackage{cleveref}

\theoremstyle{definition}

\theoremstyle{remark}

\title{Prism-Reranker: Beyond Relevance Scoring ---\\
Jointly Producing Contributions and Evidence for Agentic Retrieval}
\author{%
  DunZhang\thanks{Independent researcher. Correspondence: \href{mailto:dunnzhang0@gmail.com}{dunnzhang0@gmail.com}.}
}
\date{April 26, 2026}

\begin{document}
\maketitle

\begin{abstract}
Modern retrieval pipelines increasingly serve downstream consumers---retrieval-augmented generation (RAG)~\citep{lewis2020rag} and autonomous agents~\citep{yao2023react}---that need more than a scalar relevance score. A reranker that only tells the caller ``how relevant'' forces the agent to dump entire documents into the language-model context, wasting tokens on tangential passages, boilerplate from web crawls, and redundant background, while providing no actionable signal to drive the next planning step. We introduce \textbf{Prism-Reranker}, a family of reranker models built on the \textsc{Qwen3.5}~\citep{qwen3.5} backbone at four sizes (0.8B, 2B, 4B, 9B) that goes beyond scalar scoring. In addition to the standard \texttt{yes}/\texttt{no} relevance judgement, whenever the verdict is \texttt{yes} the model emits (i) a \emph{contribution} statement that summarizes how the document helps the query, and (ii) an \emph{evidence} passage---a self-contained rewrite that preserves every query-relevant signal from the original document while discarding redundancy and noise. Prism-Reranker is trained with a hybrid objective combining point-wise distillation from a strong commercial reranker API with supervised fine-tuning on contribution and evidence targets. We curate training data by drawing on the open-source retrieval-data aggregation released by KaLM-Embedding~\citep{hu2025kalmembedding,zhao2025kalmembeddingv2}, augmenting it with real web documents retrieved via commercial search APIs for open-domain queries and their LLM-synthesized variants, and rewriting a portion of queries into keyword-style reformulations so the model adapts to traffic issued by agents. To reconcile inconsistent labeling conventions across open corpora and to obtain crisp binary labels for the contribution-and-evidence supervision branch, we relabel data with an LLM-as-Judge ensemble that aggregates votes from five diverse frontier LLMs. On a QA subset of BEIR and on an LLM-judged evaluation of contribution and evidence quality, Prism-Reranker attains solid results across all four model sizes. We further show that the same recipe can extend existing LLM-based rerankers---augmenting Qwen3-Reranker-4B with contribution and evidence capabilities while improving its average BEIR-QA NDCG@10 by $+1.54$ over the base model. Model weights, the full training recipe, and the evaluation suite are released to the community.
\end{abstract}

\section{Introduction}
\label{sec:intro}

Neural retrieval pipelines have long followed a two-stage recipe: a fast first-stage retriever~\citep{karpukhin2020dpr,wang2022e5,chen2024bgem3} fetches a candidate pool, after which a cross-encoder reranker~\citep{nogueira2019bertrerank,xiao2023bge} reorders it. In the last two years the downstream consumer of that pipeline has changed qualitatively. Retrieval-augmented generation~\citep{lewis2020rag} and, more recently, tool-using agents~\citep{yao2023react,schick2023toolformer} now stand between retrieval and the end user: instead of a human eyeballing the top-10 snippets, a large language model reads the ranked list and either generates an answer, plans another action, or issues a follow-up query. This shift reshapes what a reranker should output.

Contemporary open-source and commercial rerankers~\citep{xiao2023bge,chen2024bgem3,zhang2025qwen3emb,wang2025jinarerankerv3} return a single scalar per \mbox{(query,~document)} pair. That output is sufficient for pure ranking, but it leaves three pain points for RAG and agent workloads. \textbf{(1) Long documents waste context.} Once a document is deemed relevant, current practice is to feed its full text into the downstream LLM, even though most passages inside a long document are tangential to the specific query. \textbf{(2) Web-crawled documents are noisy.} When agents search the open web via services such as Tavily or Exa~\citep{tavily,exa}, returned pages mix the answer with navigation menus, advertising copy, repeated disclaimers, and unrelated context; a scalar score cannot separate signal from noise. \textbf{(3) A scalar offers no planning signal.} An agent deciding whether to stop retrieving, to refine the query, or to pivot to a different tool benefits from knowing \emph{what} each candidate contributes, not merely \emph{whether} it is relevant.

We address these pain points with \textbf{Prism-Reranker}, a reranker family that jointly emits three outputs in a single forward pass. Built on the \textsc{Qwen3.5}~\citep{qwen3.5} backbone and released at 0.8B, 2B, 4B, and 9B sizes, the model (a) produces a \texttt{yes}/\texttt{no} token from which a calibrated relevance score is recovered as $\sigma(\ell_{\texttt{yes}} - \ell_{\texttt{no}})$, (b) when the verdict is \texttt{yes}, generates a short \emph{contribution} sentence describing what the document adds to the query, and (c) generates a self-contained \emph{evidence} passage that is a faithful, redundancy-stripped rewrite of the query-relevant portion of the source. The evidence field is designed to completely replace the original document when the downstream LLM consumes it, cutting context length while preserving answerable content; the contribution field is designed to be read at the agent-planning layer.

Realizing this interface requires rethinking both training and data. On the training side, Prism-Reranker is supervised by two complementary signals combined through a single loss: a point-wise distillation term that aligns $\sigma(\ell_{\texttt{yes}} - \ell_{\texttt{no}})$ with the teacher score from a strong, widely used commercial rerank API, and a supervised fine-tuning term on the concatenated \texttt{yes}/\texttt{no}, contribution, and evidence targets. A small ablation confirms that the simplest point-wise distillation suffices for the ranking head---listwise or pairwise objectives did not help, which we attribute to the inherent fitting capacity of a cross-encoder scorer.

On the data side, we observe that training-time query distribution is the single largest factor affecting how Prism-Reranker behaves in production. We therefore: \textbf{(i)} build on the open-source retrieval-data aggregation released by KaLM-Embedding~\citep{hu2025kalmembedding,zhao2025kalmembeddingv2}, which itself unifies a large pool of English, Chinese, and multilingual retrieval corpora, and uniformly sample across its constituent datasets; \textbf{(ii)} take open-source queries together with LLM-synthesized queries and feed them into commercial web search APIs (Tavily and Exa) to collect real web documents, producing a data slice whose surface statistics match what an agent actually sees at inference; \textbf{(iii)} rewrite roughly 30\% of queries into short keyword-style strings through a dedicated \textsc{DeepSeek-V3.2} rewriting pass, because agent-issued queries are often keyword bags rather than well-formed questions; \textbf{(iv)} balance the final dataset along both document length and teacher score so that no length~$\times$~score cell dominates.

The commercial teacher provides only a continuous score, not a binary verdict that contribution/evidence generation can be gated on. Open corpora carry their own pathology: different datasets use inconsistent criteria for what counts as ``relevant.'' We therefore deploy an \textbf{LLM-as-Judge ensemble} for binary labeling. We surveyed a large panel of frontier LLMs, measured their pairwise agreement, and deliberately selected five models whose verdicts are individually strong yet, \emph{within the candidate pool}, mutually as decorrelated as possible---\textsc{DeepSeek-V3.2}, \textsc{Qwen3.5-397B-A17B}, \textsc{Gemini-3-Flash}, \textsc{Claude-Haiku-4.5}, and \textsc{GPT-5.4-mini}---and declare a pair relevant when at least three of the five agree. We are explicit that the resulting pairwise agreements remain high in absolute terms (\cref{sec:data:judge}); they represent the maximum diversity attainable from a pool of frontier judges that already largely concur on relevance, not low correlation in the Landis--Koch sense. This procedure reconciles heterogeneous dataset conventions behind a single crisp definition and supplies the binary positive/negative tag that determines whether a sample carries a full \texttt{<contribution>}/\texttt{<evidence>} target text or only the single token \texttt{no} during supervised fine-tuning.

Evaluation is reported on two axes. For ranking, we adapt the BEIR~\citep{thakur2021beir} evaluation protocol used by \citet{wang2025jinarerankerv3} and report NDCG@10 on the QA-style subset of BEIR (dropping tasks whose semantics are not question answering). For contribution and evidence quality---a capability for which no standard benchmark exists---we design a multi-dimensional, \textsc{DeepSeek-V4-Pro}-judged evaluation that scores faithfulness, completeness, redundancy, and contribution specificity. Prism-Reranker achieves solid results on both axes at every released size.

\paragraph{Contributions.} In summary:
\begin{itemize}
    \item We introduce \textbf{Prism-Reranker}, the first open reranker family, to the best of our knowledge, that jointly outputs a calibrated relevance score, a contribution statement, and a self-contained evidence passage in a single forward pass---an interface targeted at RAG and agent workloads rather than human-facing search.
    \item We present a \textbf{hybrid distillation-plus-SFT training recipe} in which a single combined loss applies point-wise teacher-score regression and structured-text supervised fine-tuning to every training sample; the only difference between positive and negative samples is the SFT target text (full \texttt{<contribution>}/\texttt{<evidence>} target vs.\ a single \texttt{no} token).
    \item We release a \textbf{data pipeline} that combines uniform sampling from KaLM-Embedding's open-source aggregation of retrieval corpora, agent-realistic web documents retrieved through commercial search APIs, keyword-style query reformulation, and length~$\times$~score balanced curation.
    \item We show that a \textbf{5-judge LLM-as-Judge ensemble}, drawn by maximum-disagreement selection from a pool of frontier judges that already largely concur on relevance (so the resulting panel is more diverse than an ad-hoc choice but still highly correlated in absolute terms; \cref{sec:data:judge}), reconciles heterogeneous labeling conventions across open corpora into a single crisp positive/negative tag that determines which samples carry contribution-and-evidence supervision.
    \item We demonstrate a \textbf{flexible extension variant} that augments an existing LLM-based reranker (\textsc{Qwen3-Reranker-4B}) with contribution and evidence capabilities via self-distillation, lifting its average BEIR-QA NDCG@10 by $+1.54$ over the base model without invoking a commercial teacher.
    \item We release five models---four sizes (0.8B, 2B, 4B, 9B) on the Qwen3.5 backbone plus Prism-Reranker-4B-exp, an experimental extension of Qwen3-Reranker-4B---along with the full training code, the balanced training corpus, and the evaluation harness for both ranking and contribution/evidence quality.
\end{itemize}

\section{Related Work}
\label{sec:related}

\paragraph{Cross-encoder and generative rerankers.}
Cross-encoder rerankers were established by \citet{nogueira2019bertrerank} and \citet{nogueira2020monot5}, and remain the workhorse of modern retrieval pipelines through families such as BGE~\citep{xiao2023bge,chen2024bgem3}, Jina~\citep{wang2025jinarerankerv3}, and Qwen3-Reranker~\citep{zhang2025qwen3emb}. A second line of work prompts or fine-tunes large language models to rank candidates \emph{listwise}: RankGPT~\citep{sun2023rankgpt} establishes the prompting recipe, while RankVicuna~\citep{pradeep2023rankvicuna}, RankZephyr~\citep{pradeep2023rankzephyr}, and RankLLaMA~\citep{ma2024rankllama} train open models for the same task; FIRST~\citep{reddy2024first} accelerates listwise inference by reading only the first decoded token, and setwise/pairwise prompting variants~\citep{zhuang2024setwise} explore the design space further. All of these systems---whether cross-encoders or LLM rankers---return only a relevance score (or, equivalently, a permutation), which is precisely the interface our work argues is insufficient for downstream RAG and agent consumers.

\paragraph{Distillation for ranking.}
Knowledge distillation from a strong cross-encoder teacher into a lightweight student is a standard recipe both for compressing rerankers and for bootstrapping bi-encoders, with margin-MSE~\citep{hofstatter2020marginmse} and KL divergence over teacher scores both widely used; RankT5~\citep{zhuang2023rankt5} specifically distills listwise teacher signals into a sequence-to-sequence reranker. Our distillation term is deliberately the simplest possible point-wise MSE between $\sigma(\ell_{\texttt{yes}} - \ell_{\texttt{no}})$ and the commercial teacher's score. An ablation in \cref{sec:experiments} shows that listwise or pairwise distillation does not help our cross-encoder student, consistent with the observation that a high-capacity cross-encoder absorbs the teacher's preferences from point-wise targets alone.

\paragraph{Context compression and evidence selection for RAG.}
The need to shorten retrieved context before it reaches the generator has produced an active line of work that is the closest neighbor to our \emph{evidence} output. Token-level pruning approaches such as LLMLingua~\citep{jiang2023llmlingua} and LongLLMLingua~\citep{jiang2024longllmlingua} drop low-information tokens with a small auxiliary model, and LLMLingua-2~\citep{pan2024llmlingua2} learns the pruner from data; RECOMP~\citep{xu2024recomp} trains both extractive and abstractive compressors specifically for RAG, FILCO~\citep{wang2023filco} learns to filter sentences before generation, and EXIT~\citep{hwang2024exit} performs context-aware extractive compression at retrieval time. CompAct~\citep{yoon2024compact} compresses iteratively, while xRAG~\citep{cheng2024xrag} pushes compression all the way to a single soft token. Self-RAG~\citep{asai2024selfrag} interleaves retrieval with self-emitted critique tokens that judge support and utility. Two structural differences separate Prism-Reranker from this body of work. First, every method above is a \emph{post-retrieval} module that runs after a separate reranker, adding another inference call to the pipeline; we fold compression into the reranker so a single forward pass yields both the relevance verdict and the compressed evidence. Second, those methods emit compressed text unconditionally on every candidate; we generate evidence only when the \texttt{yes}/\texttt{no} head fires, so tail-of-list and irrelevant documents incur no generation cost at all. The \emph{contribution} field is, to the best of our knowledge, without direct precedent in the reranker literature: it is an explicit planning signal targeted at the agent layer rather than at the answer-generation layer.

\paragraph{LLM-as-Judge for relevance labeling.}
Using strong LLMs as cheap stand-ins for human judges has become standard practice since \citet{zheng2023llmasajudge,zhu2023judgelm,wang2024pandalm}, and the practice has been imported into IR for relevance assessment by \citet{faggioli2023perspectives}. \citet{verga2024poll} (``Replacing Judges with Juries'') further argue that an ensemble of smaller, diverse judges can match or beat a single frontier judge while being cheaper and less biased. Our LLM-as-Judge setup sits in this lineage but pushes the diversity argument operationally: rather than picking judges by capability alone, we measure pairwise agreement across a large panel of frontier models and deliberately select the five whose verdicts are individually strong yet, \emph{relative to the rest of the pool}, the most mutually decorrelated, then take a 3-of-5 majority. We do not claim the resulting panel is decorrelated in an absolute Landis--Koch sense (\cref{sec:data:judge})---it is not; we claim only that we have removed the most redundant judges from a pool that would otherwise cluster more tightly. This produces a single binary label that is consistent across heterogeneous source corpora and supplies the positive/negative tag that gates which samples carry full \texttt{<contribution>}/\texttt{<evidence>} SFT supervision (\cref{sec:method:train}).

\section{Method}
\label{sec:method}

This section defines the model interface (\cref{sec:method:arch}) and the training objective (\cref{sec:method:train}). Implementation details---LoRA configuration, optimizer, and loss weights---are deferred to \cref{sec:experiments}.

\subsection{Model}
\label{sec:method:arch}

\paragraph{Backbone.}
Prism-Reranker uses the \textsc{Qwen3.5}~\citep{qwen3.5} causal LM as its backbone, at four sizes (0.8B, 2B, 4B, 9B), with no architectural modification. All four sizes share a single training recipe.

\paragraph{Input format.}
Each forward pass takes one $(\text{query}, \text{document})$ pair, formatted with a fixed raw prompt template that ends in an empty \texttt{<think></think>} block placed inside the assistant turn (full template in \cref{app:prompt}). We deliberately use the raw template rather than \texttt{apply\_chat\_template()} because both the position of the first decoded token and the prompt boundary used for relevance scoring (below) must be deterministic across samples.

\paragraph{Output and relevance score.}
The model is trained so that the very first decoded token is either \texttt{yes} or \texttt{no}, followed---when the verdict is \texttt{yes}---by an XML-tagged \texttt{<contribution>}\,\ldots\,\texttt{</contribution>} sentence and an \texttt{<evidence>}\,\ldots\,\texttt{</evidence>} passage. When the verdict is \texttt{no}, the model is trained to stop after the label, so irrelevant documents incur no generation cost beyond a single token.

\begin{figure}[t]
  \centering
  \includegraphics[width=0.92\linewidth]{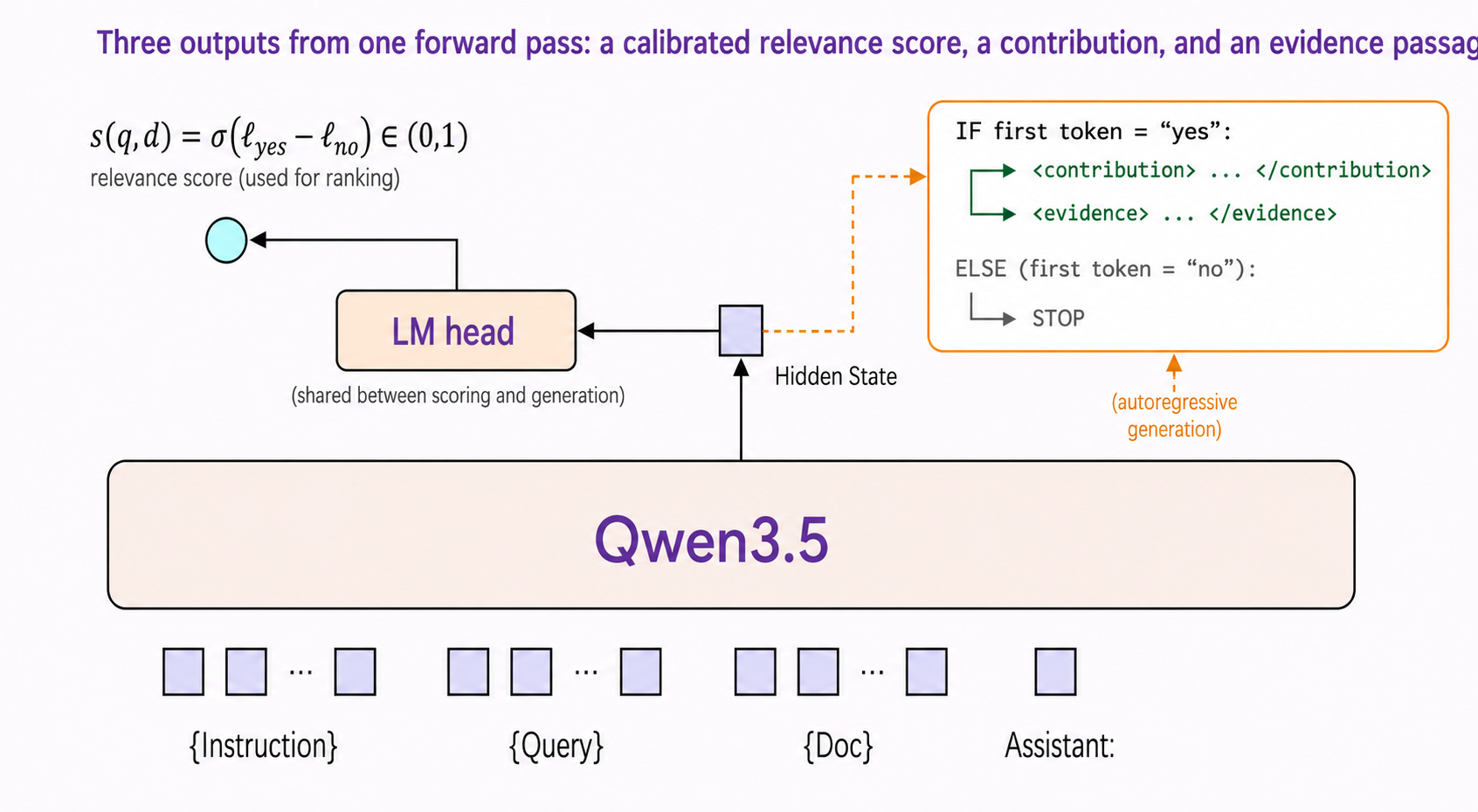}
  \caption{Three outputs from one forward pass. The same LM head is reused for (i) reading the calibrated relevance score $s(q,d) = \sigma(\ell_{\texttt{yes}} - \ell_{\texttt{no}})$ at the prompt boundary, and (ii) autoregressively generating the \texttt{<contribution>} and \texttt{<evidence>} fields when the first decoded token is \texttt{yes}. Irrelevant pairs stop after the single \texttt{no} token, so generation cost is paid only on positives.}
  \label{fig:arch}
\end{figure}

The relevance score is read directly from the logits at the prompt boundary. Let $\ell_{\texttt{yes}}$ and $\ell_{\texttt{no}}$ denote the logits at the position whose next-token prediction is the label token. We define
\begin{equation}
s(q, d) \;=\; \sigma\!\left(\ell_{\texttt{yes}} - \ell_{\texttt{no}}\right),
\label{eq:score}
\end{equation}
which is a calibrated probability in $(0, 1)$ used for ranking. \cref{fig:arch} summarizes the resulting interface: the score is read from the LM head at the prompt boundary, and the same head autoregressively continues to emit the contribution and evidence fields whenever the verdict is \texttt{yes}.

\subsection{Training Objective}
\label{sec:method:train}

We combine point-wise distillation against a commercial teacher reranker with supervised fine-tuning on structured text targets. Both losses are applied to every training sample and share one forward pass.

\paragraph{Two sample types, one objective.}
Training data has just two types of pairs, distinguished by the binary verdict produced by the LLM-as-Judge ensemble of \cref{sec:data:judge}: \emph{positive} pairs (ensemble = \texttt{yes}) and \emph{negative} pairs (ensemble = \texttt{no}). Both types are supervised by the same total loss
\begin{equation}
\mathcal{L} \;=\; \gamma_{\text{point}}\,\mathcal{L}_{\text{point}} \;+\; \gamma_{\text{sft}}\,\mathcal{L}_{\text{sft}},
\label{eq:loss-total}
\end{equation}
where $\mathcal{L}_{\text{point}}$ regresses the student score $s(q,d)$ against the commercial teacher score and $\mathcal{L}_{\text{sft}}$ supervises the structured text output. The only difference between the two sample types is the SFT target text: positives carry a full target \texttt{yes\,$\Vert$\,<contribution>\ldots</contribution>\,$\Vert$\,<evidence>\ldots</evidence>}, while negatives carry the single token \texttt{no}.

\paragraph{Point-wise distillation.}
For a pair $(q, d)$ with teacher score $y(q, d) \in [0, 1]$,
\begin{equation}
\mathcal{L}_{\text{point}} \;=\; \bigl(s(q, d) - y(q, d)\bigr)^2.
\label{eq:loss-point}
\end{equation}
We use the teacher's score directly as the regression target.

\paragraph{Supervised fine-tuning.}
With target text $T$ as defined above (full \texttt{yes}-target for positives; the single token \texttt{no} for negatives),
\begin{equation}
\mathcal{L}_{\text{sft}} \;=\; -\sum_{t \in T} \log p_\theta\bigl(t \mid q, d, t_{<}\bigr),
\label{eq:loss-sft}
\end{equation}
where prompt tokens are masked out (label $=-100$). Because the first target token is always the verdict, this loss directly supervises the binary classification head; $\mathcal{L}_{\text{point}}$ adds a continuous, score-aligned signal on top.

\paragraph{Extension to existing rerankers.}
The training framework above assumes a general-purpose LLM backbone that must learn relevance scoring from scratch. However, several recent rerankers---notably \textsc{Qwen3-Reranker}~\citep{zhang2025qwen3emb}---are architecturally standard causal LMs that have already been fine-tuned for ranking but lack the ability to generate structured text beyond a relevance score. For such models the contribution and evidence capabilities can be added \emph{without} a commercial teacher. We replace the external teacher target $y(q,d)$ in \cref{eq:loss-point} with the model's own pre-training score: $y(q,d) = s_{\text{orig}}(q,d)$, where $s_{\text{orig}}$ is obtained by running the \emph{frozen} original checkpoint on each training pair before fine-tuning begins.

It is worth being precise about what each term in this extension does. The point-wise term $\mathcal{L}_{\text{point}}$ regresses the student's score against $s_{\text{orig}}$, so the loss-minimizing solution under this term alone is to reproduce the original checkpoint's score; it acts primarily as an \emph{anchor} that prevents catastrophic ranking degradation during fine-tuning. We do not claim that self-distillation can never exceed its anchor---it sometimes does in other regimes---only that, with $s_{\text{orig}}$ as the regression target, this term provides no \emph{new} information about ranking beyond what the base model already encodes. The improvement we observe over the base model is therefore most naturally attributed to the SFT branch, whose first target token is the 5-judge ensemble verdict (\cref{sec:data:judge}). The ensemble label is constructed independently of $s_{\text{orig}}$ and supplies a fresh binary supervision signal that the gate can be pulled toward when the original ranker disagrees with the ensemble. We additionally hypothesise that joint training on the structured \texttt{<contribution>}/\texttt{<evidence>} targets acts as a soft regulariser on the same hidden states that drive the relevance gate, but we have not isolated this effect with an ablation. The empirical result on \textsc{Qwen3-Reranker-4B} (\cref{sec:experiments:beir}) is consistent with this picture: ranking quality improves, and the improvement is most plausibly driven by the ensemble-label SFT signal rather than by self-distillation alone.

The rest of the training recipe---loss form, LoRA configuration, optimizer schedule---remains unchanged. We demonstrate this variant with \textsc{Qwen3-Reranker-4B} in \cref{sec:experiments}.

\section{Data}
\label{sec:data}

\begin{figure}[t]
  \centering
  \includegraphics[width=\linewidth]{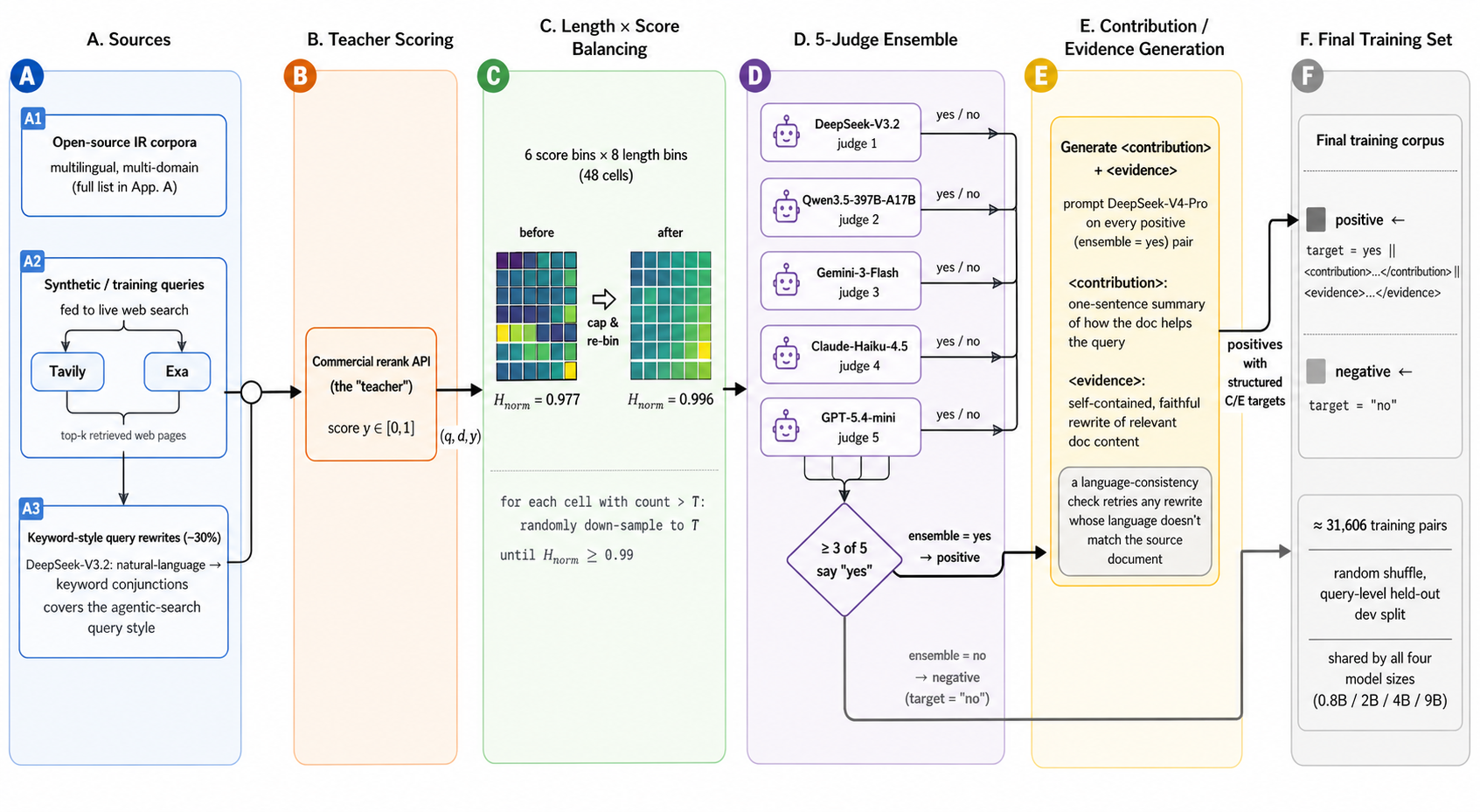}
  \caption{Data construction pipeline. Open-source IR corpora and live web pages retrieved through Tavily and Exa---for both natural-language and keyword-style queries---are scored by a commercial rerank API (the teacher) and balanced across a $6 \times 8$ length-by-score grid. A 5-judge LLM ensemble then issues a binary positive/negative tag by 3-of-5 majority vote; positives receive structured \texttt{<contribution>}/\texttt{<evidence>} targets generated by \textsc{DeepSeek-V4-Pro}, while negatives carry the single token \texttt{no} as the SFT target. The teacher score serves as the point-wise regression target on every pair; the binary tag determines only which SFT target text is used.}
  \label{fig:pipeline}
\end{figure}

Training data is assembled through a multi-stage pipeline that combines broad coverage (diverse corpora and live web documents) with high-quality binary labels and structured text supervision. \cref{fig:pipeline} gives a high-level overview; the remainder of this section walks through each stage.

\subsection{Data Sources}
\label{sec:data:sources}

\paragraph{Open-source corpora.}
Rather than reassembling a heterogeneous collection of public retrieval datasets ourselves, we build on the open-source retrieval-data aggregation released by KaLM-Embedding~\citep{hu2025kalmembedding,zhao2025kalmembeddingv2,kalm2025dataset}. Their release consolidates a large pool of English, Chinese, and multilingual retrieval corpora---spanning open-domain QA (MS MARCO, NQ, HotpotQA, TriviaQA), domain-specific QA (PubMedQA, FiQA, LegalQA), e-commerce (ESCI), web search (DuReader, mMARCO, T2Ranking), and multilingual benchmarks (MIRACL, Mr.~TyDi)---into a single uniform format. We uniformly sample across its constituent datasets so that no single domain or language dominates the training distribution.

\paragraph{Web-sourced documents.}
Curated benchmarks tend to be cleaner than documents encountered in production. To close this gap, we issue training queries against two live web-search APIs---Tavily~\citep{tavily} and Exa~\citep{exa}---and retain the top-retrieved pages as additional training documents. Web-sourced data better reflects the heterogeneous, often noisy content that real-world agentic retrieval systems must handle.

\paragraph{Keyword query conversion.}
Agentic search pipelines frequently emit keyword conjunctions rather than full natural-language questions. To improve robustness to this query style, approximately 30\% of training queries are rewritten into keyword form using DeepSeek-V3.2\footnote{All DeepSeek models used throughout this work (V3.2, V4-Flash, and V4-Pro) are invoked without reasoning/thinking mode.} before retrieval and scoring, producing additional training pairs from the same document pool.

\subsection{Teacher Scoring}
\label{sec:data:teacher}

Each query--document pair is scored by a strong commercial rerank API (the \emph{teacher}). The score $y \in [0, 1]$ returned by the API is used directly as the point-wise regression target for $\mathcal{L}_{\text{point}}$ (\cref{sec:method:train}).

We deliberately do not name the teacher API in this paper. As a reproducibility mitigation, we release the cached teacher scores for every training pair together with the model weights, so that downstream users can recover the same training targets without invoking the teacher API themselves.

\subsection{LLM-as-Judge Annotation}
\label{sec:data:judge}

Teacher scores are continuous and carry no natural decision threshold; furthermore, relevance criteria differ across open-source datasets, making their binary labels inconsistent with each other. We therefore augment the teacher signal with ensemble-voted binary labels from a panel of frontier LLMs~\citep{zheng2023llmasajudge,verga2024poll}.

\begin{figure}[t]
  \centering
  \includegraphics[width=\linewidth]{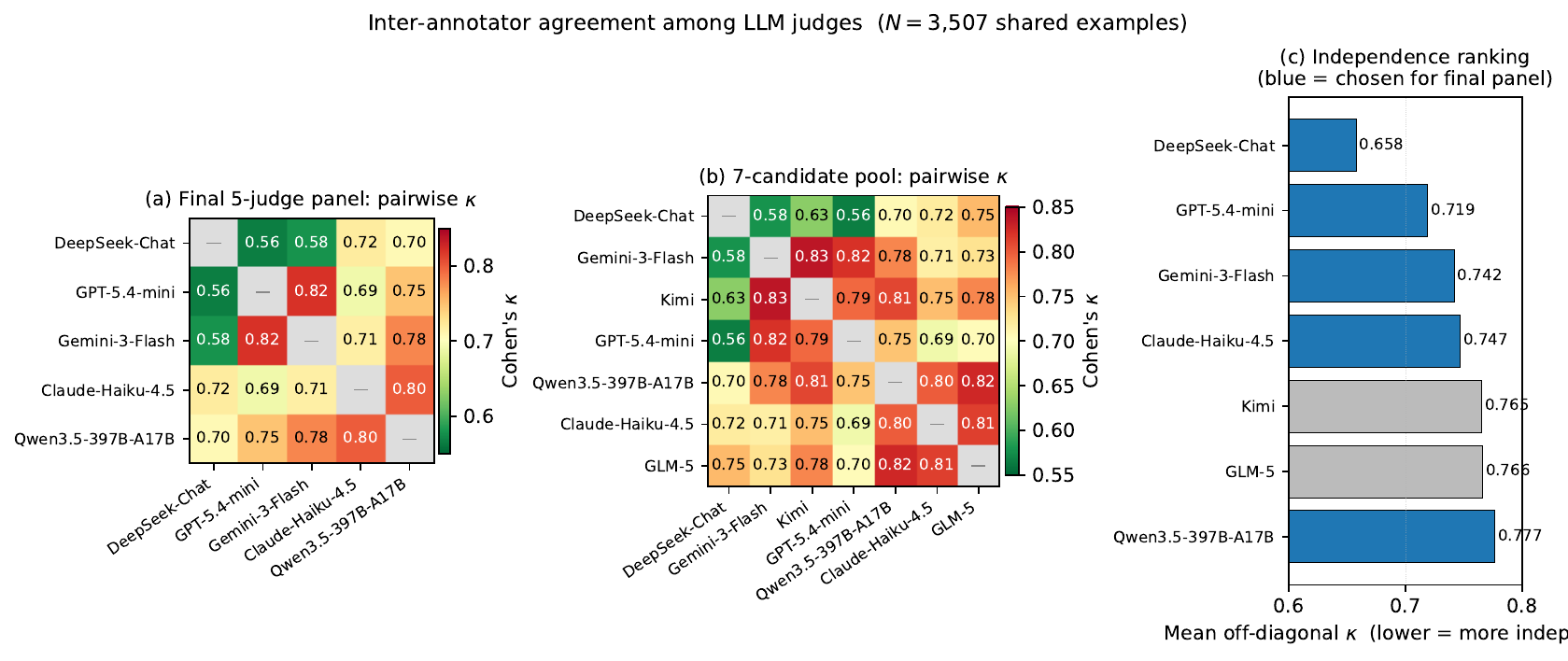}
  \caption{Inter-annotator agreement among LLM judges, measured as Cohen's $\kappa$ on $N{=}3{,}507$ shared examples. \textbf{(a)} Pairwise $\kappa$ over the final 5-judge panel; the largest pairwise $\kappa$ is $0.82$. By Landis--Koch this is still ``almost perfect'' agreement in absolute terms---the panel is the \emph{relatively} most decorrelated 5-subset of the candidate pool, not a decorrelated panel in any absolute sense. \textbf{(b)} Pairwise $\kappa$ over the 7-model candidate pool. \textbf{(c)} Independence ranking by mean off-diagonal $\kappa$ (lower is more relatively independent within this pool); blue bars mark the five judges retained for the ensemble, gray bars mark the two candidates dropped because they cluster too tightly with judges already selected.}
  \label{fig:judge_kappa}
\end{figure}

After surveying a wide range of models and computing pairwise inter-annotator agreement, we select five judges whose verdicts are, \emph{within the surveyed pool}, mutually as decorrelated as we could make them while keeping individual judge quality high: DeepSeek-V3.2, Qwen3.5-397B-A17B, Gemini-3-Flash, Claude-Haiku-4.5, and GPT-5.4-mini. The largest pairwise Cohen's $\kappa$ among the panel is $0.82$, and the smallest is also high---we acknowledge that on the Landis--Koch scale these values still indicate substantial-to-almost-perfect agreement. The point of the selection is therefore not that the panel is decorrelated in absolute terms (it is not; frontier LLMs largely concur on relevance), but that we have removed the most redundant judges from a pool that would otherwise cluster even more tightly, so that the 3-of-5 majority vote draws on the maximum diversity actually available. \cref{fig:judge_kappa} reports the full pairwise agreement matrix over both the final 5-judge panel and the broader candidate pool from which it was selected. Each judge receives a unified, unambiguous relevance rubric and returns a binary yes/no verdict; the ensemble label is decided by a 3-of-5 majority vote.

The ensemble verdict and the teacher score are used independently: the verdict assigns each pair to the positive or negative bucket and thereby chooses which SFT target text is emitted, while the teacher score is regressed on every pair (\cref{sec:method:train}).

\subsection{Contribution and Evidence Generation}
\label{sec:data:ce}

For every pair that the ensemble votes \texttt{yes} (positives), we prompt \textsc{DeepSeek-V4-Pro} to generate the structured \texttt{<contribution>} and \texttt{<evidence>} targets consumed by $\mathcal{L}_{\text{sft}}$. For pairs voted \texttt{no} (negatives), the SFT target is simply the single token \texttt{no}, with no continuation.

\subsection{Length--Score Balancing}
\label{sec:data:balance}

\begin{figure}[t]
  \centering
  \includegraphics[width=\linewidth]{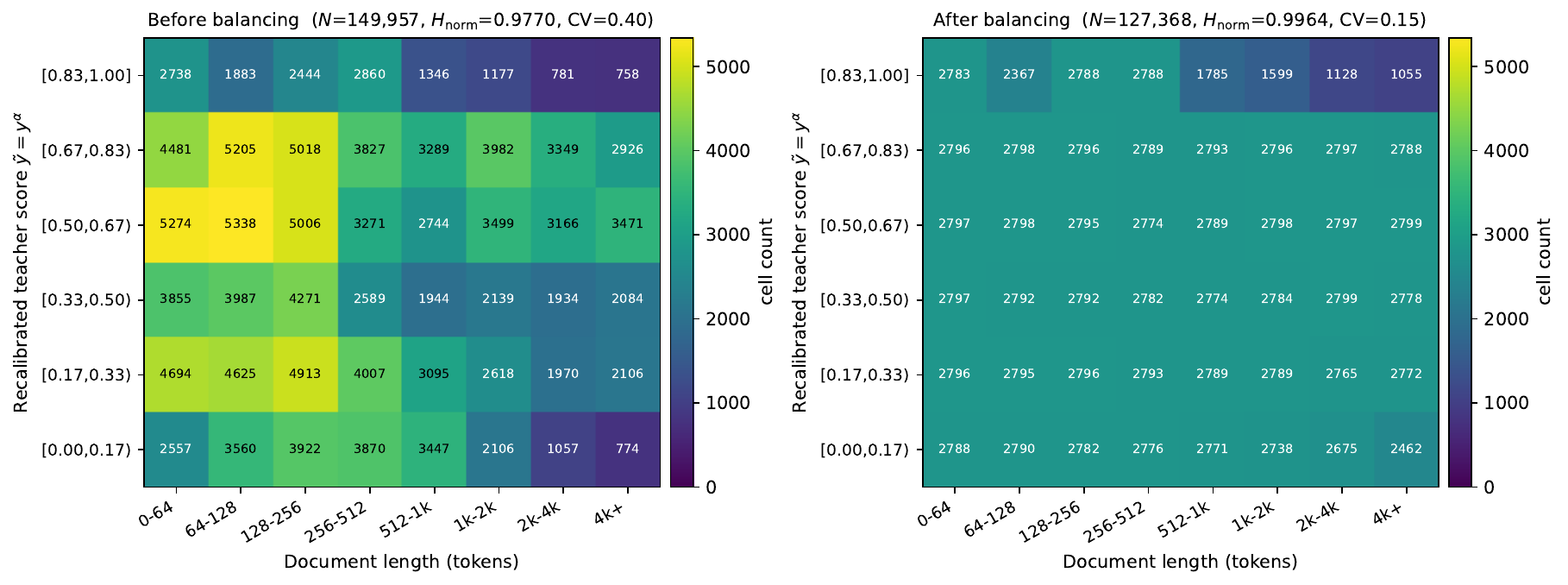}
  \caption{Joint distribution of training pairs over teacher score $y$ (rows, six equal-width bins) and document token length (columns, eight log-spaced bins). \textbf{Before balancing}, short and high-scoring documents dominate ($H_{\text{norm}} = 0.977$, CV $= 0.40$). \textbf{After balancing}, every cell is capped, raising entropy to $H_{\text{norm}} = 0.996$ (CV $= 0.15$) while retaining $\sim 85\%$ of the original pairs.}
  \label{fig:balance}
\end{figure}

Raw web-search data is heavily skewed: short, high-relevance documents are overrepresented relative to long or marginally-relevant ones. We measure distributional uniformity by placing each training example in one of $6 \times 8 = 48$ cells defined by six equal-width bins over the teacher score $y$ and eight log-spaced bins over document token count ($[0,64)$, $[64,128)$, $\ldots$, $[4096,+\infty)$). Uniformity is quantified by the normalized cell entropy
\begin{equation}
H_{\text{norm}} = -\frac{1}{\ln 48}\sum_{i=1}^{48} p_i \ln p_i,
\label{eq:entropy}
\end{equation}
where $p_i$ is the fraction of samples in cell $i$. We cap over-populated cells via random under-sampling until $H_{\text{norm}} \ge 0.99$, retaining the majority of data while achieving a near-uniform joint distribution across score and length. \cref{fig:balance} contrasts the cell counts before and after balancing on the web-search slice of our corpus: $H_{\text{norm}}$ rises from $0.977$ to $0.996$ while $85\%$ of the original samples are kept.

The final corpus supports sequences up to 10,240 tokens and covers multiple languages, with English and Chinese as the primary languages and smaller proportions of other languages.

\section{Experiments}
\label{sec:experiments}

\subsection{Training Setup}
\label{sec:experiments:setup}

All four Qwen3.5-based model sizes (0.8B, 2B, 4B, 9B) share a single training recipe. We apply LoRA to all attention projections---including the linear-attention modules \texttt{in\_proj\_\{qkv,a,b,z\}}---and all MLP layers. The rank is $r{=}64$, $\alpha{=}128$ for the three smaller sizes and $r{=}32$, $\alpha{=}64$ for 9B, with no dropout. Optimization uses AdamW with learning rate $10^{-5}$, weight decay 0.01, 100-step linear warm-up, and cosine decay. Sequences are truncated to 10,240 tokens. Per-device batch size is 1 with gradient accumulation over 8 steps; training runs for 2 epochs (3 epochs for 0.8B). Loss weights are $\gamma_{\text{point}} = 20$ and $\gamma_{\text{sft}} = 1.0$. Training was carried out on a single NVIDIA RTX 4090 (24~GB) for the 0.8B, 2B, and 4B sizes, and on a single NVIDIA A800 (80~GB) for the 9B size. All released checkpoints are distributed as fully merged weights, not LoRA adapters.

\paragraph{Released artifacts.}
All five checkpoints---the four Qwen3.5-based sizes and the Qwen3-Reranker-4B-based extension variant---are publicly released on the Hugging Face Hub:
\begin{itemize}
\setlength{\itemsep}{0pt}
\item Prism-Qwen3.5-Reranker-0.8B: \url{https://huggingface.co/infgrad/Prism-Qwen3.5-Reranker-0.8B}
\item Prism-Qwen3.5-Reranker-2B: \url{https://huggingface.co/infgrad/Prism-Qwen3.5-Reranker-2B}
\item Prism-Qwen3.5-Reranker-4B: \url{https://huggingface.co/infgrad/Prism-Qwen3.5-Reranker-4B}
\item Prism-Qwen3.5-Reranker-9B: \url{https://huggingface.co/infgrad/Prism-Qwen3.5-Reranker-9B}
\item Prism-Qwen3-Reranker-4B-exp: \url{https://huggingface.co/infgrad/Prism-Qwen3-Reranker-4B-exp}
\end{itemize}

\paragraph{Extension experiment.}
We additionally train \textbf{Prism-Reranker-4B-exp}, which applies the extension variant described in \cref{sec:method:train} to \textsc{Qwen3-Reranker-4B}~\citep{zhang2025qwen3emb}---an existing reranker whose architecture is a standard causal LM. Because this model already possesses strong ranking ability, the point-wise distillation target is its own pre-training score rather than a commercial teacher's. All other hyperparameters (LoRA rank, learning rate, loss weights, etc.) match the 4B Qwen3.5 configuration above.

\paragraph{Checkpoint selection.}
For each trained model, we evaluate all saved checkpoints on a held-out dev set using four metrics: Pearson correlation with the teacher score, Pearson correlation with the ensemble binary labels from \cref{sec:data:judge}, AUC, and accuracy at threshold $0.5$. We select the checkpoint that achieves the best trade-off across the label-driven metrics (label-Pearson, AUC, accuracy) rather than relying on training loss alone: we observed that \texttt{eval\_loss} can continue to decrease after downstream metrics have plateaued or begun to degrade---a common signature of the model over-fitting to the teacher's distribution rather than improving on the true label distribution. Under this protocol, the released checkpoints correspond to 48{,}000 training samples for the 0.8B model, one full pass over the 31{,}606-sample training set for the 2B and 4B models, and only 22{,}000 samples for the 9B model, which converges visibly faster than the smaller variants and starts to over-fit before a full epoch is completed.

\paragraph{Note on size scaling.} Because checkpoint selection optimizes dev-set quality rather than equal sample budget, the released sizes have seen different numbers of training samples (smallest seeing the most, largest the fewest). We report these counts for transparency. Cross-size NDCG@10 numbers in \cref{sec:experiments:beir} should therefore be read as the released-checkpoint Pareto frontier, not as a controlled scaling study at fixed compute or data budget; we have not run an iso-sample sweep across the four sizes and do not claim a scaling law from this table.

\subsection{Relevance Ranking on BEIR}
\label{sec:experiments:beir}

We evaluate ranking quality on a 9-dataset subset of BEIR~\citep{thakur2021beir}.
We adopt the same first-stage retrieval pipeline as Jina-Reranker-v3~\citep{wang2025jinarerankerv3}: jina-embeddings-v3 retrieves the top-100 candidates, which every reranker then re-scores.
Sharing this identical candidate set ensures that scores are directly comparable across models.
We exclude four BEIR datasets that lack a clear question--answer structure (ArguAna, FEVER, ClimateFEVER, Quora) and report NDCG@10 on the remaining nine.

\begin{table}[t]
\centering
\caption{NDCG@10 (\%) on 9 BEIR datasets. All rerankers operate on the same top-100 candidates from jina-embeddings-v3.
Baseline numbers are taken from~\citep{wang2025jinarerankerv3}; Avg.\ is computed over the 9 datasets listed here.
For jina-reranker-v3, we report the random-ordering variant (R), which is its best configuration.
$^{\ddagger}$The commercial teacher API was evaluated on 7 of the 9 datasets only; NQ and HotpotQA were skipped due to query-volume cost. We compare the teacher to our students on the shared 7-dataset subset in the text below; the 9-dataset Avg.\ column is left blank for this row to avoid a misleading comparison.
NFC\,=\,NFCorpus, SF\,=\,SciFact, SD\,=\,SCIDOCS, FQA\,=\,FiQA,
TC\,=\,TREC-COVID, TCH\,=\,Touch\'e, DBP\,=\,DBPedia,
NQ\,=\,Natural Questions, HQA\,=\,HotpotQA.}
\label{tab:beir}
\scriptsize
\setlength{\tabcolsep}{3pt}
\begin{tabular}{l c *{10}{c}}
\toprule
Model & Size & Avg. & NFC & SF & SD & FQA & TC & TCH & DBP & NQ & HQA \\
\midrule
\multicolumn{12}{l}{\textit{First-stage retriever}} \\
jina-embeddings-v3         & 0.5B & 50.09 & 36.65 & 72.40 & 19.91 & 47.47 & 77.81 & 26.55 & 41.07 & 64.31 & 64.63 \\
\midrule
\multicolumn{12}{l}{\textit{Existing rerankers}} \\
jina-reranker-v2           & 0.3B & 53.83 & 37.17 & 76.50 & 20.03 & 46.48 & 80.53 & 32.35 & 47.81 & 67.39 & 76.17 \\
mxbai-rerank-base-v2       & 0.5B & 54.14 & 37.57 & 77.76 & 18.09 & 47.33 & 82.75 & 30.71 & 48.00 & 67.74 & 77.35 \\
jina-colbert-v2            & 0.6B & 51.95 & 35.88 & 70.13 & 19.40 & 43.62 & 81.94 & 29.11 & 47.14 & 66.01 & 74.36 \\
bge-reranker-v2-m3         & 0.6B & 53.29 & 34.33 & 72.64 & 17.79 & 45.45 & 82.19 & 33.12 & 46.72 & 69.52 & 77.89 \\
Qwen3-Reranker-0.6B        & 0.6B & 51.84 & 38.37 & 74.89 & 20.98 & 43.45 & 87.08 & 27.26 & 43.54 & 56.54 & 74.41 \\
jina-reranker-v3 (R)       & 0.6B & 56.43 & 38.92 & 76.84 & 24.26 & 51.81 & 86.59 & 30.12 & 48.37 & 72.90 & 78.03 \\
mxbai-rerank-large-v2      & 1.5B & 55.43 & 37.76 & 78.86 & 18.58 & 52.75 & 81.51 & 29.81 & 49.07 & 72.46 & 78.10 \\
jina-reranker-m0           & 2.4B & 56.67 & 41.03 & 79.94 & 22.91 & 51.62 & 84.17 & 31.79 & 49.34 & 72.25 & 76.99 \\
Qwen3-Reranker-4B          & 4.0B & 57.33 & 41.56 & 78.41 & 26.01 & 52.29 & 87.08 & 33.73 & 50.81 & 69.06 & 77.03 \\
\midrule
\multicolumn{12}{l}{\textit{Commercial teacher (anonymized)}} \\
Teacher API$^{\ddagger}$   & ---  & ---   & 41.95 & 79.91 & 23.53 & 60.42 & 87.23 & 30.82 & 50.74 & ---   & ---   \\
\midrule
\multicolumn{12}{l}{\textit{Prism-Reranker (ours)}} \\
Prism-Reranker-0.8B        & 0.8B & 51.92 & 39.64 & 75.52 & 19.26 & 43.23 & 82.38 & 32.79 & 44.86 & 60.52 & 69.05 \\
Prism-Reranker-2B          & 2B   & 53.32 & 40.53 & 76.56 & 19.85 & 45.89 & 84.35 & 32.34 & 46.14 & 63.67 & 70.56 \\
Prism-Reranker-4B          & 4B   & 54.18 & 41.49 & 77.03 & 20.85 & 48.91 & 83.61 & 30.33 & 47.16 & 68.64 & 69.62 \\
Prism-Reranker-9B          & 9B   & 55.02 & 41.43 & 79.90 & 21.74 & 50.28 & 83.39 & 30.57 & 47.75 & 69.16 & 70.97 \\
\midrule
\multicolumn{12}{l}{\textit{Extension experiment (ours)}} \\
Prism-Reranker-4B-exp\footnotemark & 4B & 58.87 & 42.33 & 78.39 & 26.13 & 55.05 & 89.05 & 38.55 & 51.21 & 72.67 & 76.42 \\
\bottomrule
\end{tabular}
\end{table}

\footnotetext{Based on \textsc{Qwen3-Reranker-4B} rather than Qwen3.5; trained with self-distillation to preserve the original model's ranking quality (see \cref{sec:method:train}).}

Table~\ref{tab:beir} reports the results.
Among the four Qwen3.5-based models, average NDCG@10 on the released checkpoints rises monotonically from 51.92 (0.8B) to 55.02 (9B). We caution against reading this as a clean size-scaling result: as noted in \cref{sec:experiments:setup}, checkpoint selection optimizes dev-set quality rather than equal sample budget, and the larger models in fact see fewer training samples than the smaller ones. The trend reflects the released-checkpoint Pareto frontier under our recipe, not a controlled scaling law. These models are trained from a general-purpose LLM backbone and must learn relevance scoring from scratch via distillation; unlike the baselines, they simultaneously acquire the ability to produce contribution and evidence outputs.

\paragraph{Comparison with the commercial teacher.}
On the 7-dataset subset that the teacher API could be evaluated on, the teacher achieves an average NDCG@10 of $53.51$, while Prism-Reranker-9B / 4B / 2B / 0.8B reach $50.72$ / $49.91$ / $49.38$ / $48.24$ on the \emph{same} 7 datasets. All four Qwen3.5-based students therefore land below the teacher by roughly $2.8$ to $5.3$ points. We attribute this gap to the multi-task burden carried by the students: in addition to fitting the teacher's relevance score, each student must simultaneously learn to generate the \texttt{<contribution>} and \texttt{<evidence>} fields from scratch on the same forward pass, and a non-trivial share of the model's capacity is necessarily redirected away from pure score-fitting. A student trained primarily by point-wise score regression typically tracks rather than surpasses its teacher on the teacher's own scoring distribution---improvements over the teacher are possible but not the norm; on top of this practical ceiling, the joint contribution-and-evidence objective costs measurable ranking quality on this benchmark. The relevant question for these sizes is therefore whether the structured-output capability is worth a few NDCG@10 points of ranking, not whether the students can outrank a teacher they were directly distilled from.

\paragraph{The 4B-exp result.}
A complementary picture comes from \textbf{Prism-Reranker-4B-exp}, which exercises the extension variant (\cref{sec:method:train}) by replacing the commercial teacher with \textsc{Qwen3-Reranker-4B}'s own scores as a self-distillation anchor. \textsc{Qwen3-Reranker-4B} is itself a BEIR-strong reranker (57.33 avg over 9 datasets in \cref{tab:beir}), and applying the same Prism recipe on top lifts ranking quality to 58.87---a $+$1.54 gain over its own anchor; seven of the nine datasets improve and the remaining two change by less than one point. On the 7-dataset subset that includes the teacher, 4B-exp reaches $54.39$, marginally above the commercial teacher's $53.51$ on the same 7 datasets. We do not read this as evidence that the Prism recipe \emph{outperforms} commercial rerankers in general---4B-exp is anchored to a different, BEIR-strong open-source reranker rather than to the commercial API---only that, when paired with a sufficiently strong anchor, the recipe absorbs the joint contribution-and-evidence objective without surrendering ranking quality.

It is worth being precise about where the $+$1.54 over \textsc{Qwen3-Reranker-4B} itself comes from. The point-wise self-distillation term $\mathcal{L}_{\text{point}}$ regresses against \textsc{Qwen3-Reranker-4B}'s own pre-training score, and its loss-minimizing solution is to reproduce that score; in expectation this term anchors ranking near the base model rather than driving it higher, and on its own it carries no information about ranking beyond what the base model already encodes. The observed improvement is therefore most plausibly driven by the SFT branch: the first SFT target token is the 5-judge ensemble verdict, which was constructed independently of \textsc{Qwen3-Reranker-4B} and supplies a fresh binary label whenever the original ranker disagrees with the ensemble. Joint training on the structured contribution and evidence targets plausibly contributes additional regularisation on the shared hidden states that drive the gate; we report this as a hypothesis rather than an ablated fact.

\subsection{Contribution and Evidence Quality}
\label{sec:experiments:ce}

No existing benchmark targets the quality of jointly produced contribution and evidence fields. We therefore construct a dedicated evaluation set and assess the model's structured outputs along nine complementary dimensions---three rule-based and six LLM-judged.

\paragraph{Evaluation set.}
The evaluation set is drawn from the same pipeline described in \cref{sec:data} and split at the \emph{query level}: no query appears in both the training and evaluation partitions. This strict partitioning prevents information leakage through queries that share different documents across splits---a weaker split that only separates query--document pairs could still expose the model to the same query at training time.

\paragraph{Rule-based metrics.}
\texttt{label\_match} measures binary classification accuracy: whether the first decoded token (\texttt{yes}/\texttt{no}) agrees with the ensemble ground-truth label.
\texttt{format\_score} $\in \{0, 0.4, 0.7, 1.0\}$ evaluates structural compliance under three cases. \textbf{(no)} If the first decoded token is \texttt{no} and the model emits no further text, the sample scores $1.0$; a \texttt{no} first token followed by any continuation scores $0$, since the protocol requires irrelevant pairs to terminate immediately. \textbf{(yes)} If the first token is \texttt{yes}, the sample receives $+0.4$ for the parseable verdict, plus $+0.3$ for a well-formed \texttt{<contribution>} field ($>$10~characters) and $+0.3$ for a well-formed \texttt{<evidence>} field ($>$10~characters), capped at $1.0$. \textbf{(other)} Any first token that is neither \texttt{yes} nor \texttt{no} scores $0$. The three cases are symmetric in the sense that a fully-compliant output---whether positive or negative---can attain the full $1.0$.

\paragraph{Entity fidelity.}
For samples where both the ground-truth label and the model prediction are \texttt{yes}, we extract key entities from the generated evidence via a two-pass approach: (1)~an LLM (DeepSeek-V4-Flash) extracts proper nouns, technical terms, model codes, and URLs; (2)~regex patterns capture numbers, percentages, and dates. Each extracted entity is validated by checking for a verbatim substring match in the source document. The fidelity score is the fraction of entities present in the document, directly quantifying factual hallucination.

\paragraph{LLM-as-Judge protocol.}
For the same yes/yes subset, we employ DeepSeek-V4-Pro as a single-call judge that scores six quality dimensions on an integer 1--5 scale. The judge is calibrated with a \emph{start-from-3} anchor: the default score for an acceptable sample is~3; a score of~4 requires demonstrable merit with no shortcoming, while~5 (expert-level) is reserved for outputs that surpass the source document in clarity---most samples are expected to fall in the 2--4 range. Hard disqualification rules further constrain scores when critical failures are detected; for example, hallucinated numbers force \texttt{evidence\_faithfulness}\,$=$\,1 regardless of other qualities. The six dimensions are:

\begin{itemize}
  \item \textbf{contribution\_accuracy}\,---\,Does the contribution faithfully describe what the document actually contributes to the query? Fabrication or empty boilerplate (e.g.\ ``this article discusses\ldots'') caps the score at~2.
  \item \textbf{contribution\_coverage}\,---\,Does a single sentence capture all key contribution points without omission or redundancy?
  \item \textbf{evidence\_faithfulness}\,---\,The most critical dimension. Are numbers, named entities, and hedging language (``approximately,'' ``reportedly'') preserved verbatim from the source? Any altered number or fabricated causal claim forces a score of~1.
  \item \textbf{evidence\_self\_contained}\,---\,Can the evidence alone answer the query without referring back to the original document? Unresolved pronouns (``this method,'' ``they'') or missing qualifiers (sample size, time scope) lower the score.
  \item \textbf{evidence\_concision}\,---\,Has irrelevant background been removed? Verbatim copying of the source without condensation is capped at~3.
  \item \textbf{language\_consistency}\,---\,Binary (5 or 1): the output language must match the document language; for multilingual documents, it must match the query language or default to English. Proper nouns and technical terms are excluded from this check.
\end{itemize}

\paragraph{Reading the LLM-judged columns.}
Two structural facts about \cref{tab:ce} should be flagged before the table is read. Our SFT contribution and evidence targets are themselves \textsc{DeepSeek-V4-Pro} generations (\cref{sec:data:ce}) and the judge in this section is also \textsc{DeepSeek-V4-Pro}: the six 1--5 LLM-judged columns therefore measure how closely a model reproduces the V4-Pro output style on quality dimensions that V4-Pro itself defines. Absolute LLM-judge scores should be read as fidelity-to-teacher rather than absolute quality, and cross-model comparisons within these columns are robust only when the candidates are not themselves V4-Pro family members. The metric that bypasses this loop is \textbf{label\_match} (which compares against the 5-judge ensemble verdict, constructed independently of V4-Pro). \textbf{entity\_fidelity}, although a deterministic substring check, is upper-bounded by the fidelity of the V4-Pro-generated SFT targets themselves; absolute numbers reflect how often the student copies entities verbatim, but should not be over-interpreted as ``hallucination measurement at frontier-LLM scale,'' since the teacher V4-Pro---which the student imitates---may itself paraphrase entities at some non-zero rate.

\begin{table}[t]
\centering
\caption{Contribution and evidence quality. LLM scores are 1--5; fidelity is $[0,1]$; label and format are accuracy / score in $[0,1]$.
lbl\,=\,label\_match, fmt\,=\,format\_score, fid\,=\,entity\_fidelity,
c-acc\,=\,contribution\_accuracy, c-cov\,=\,contribution\_coverage,
e-fth\,=\,evidence\_faithfulness, e-sc\,=\,evidence\_self\_contained,
e-con\,=\,evidence\_concision, lang\,=\,language\_consistency.
$^{\dagger}$DeepSeek-V4-Flash was prompted in this run to emit only contribution and evidence (no relevance gate), so the lbl/fmt columns are not measured here.}
\label{tab:ce}
\small
\setlength{\tabcolsep}{4pt}
\begin{tabular}{lcccccccccc}
\toprule
Model & lbl & fmt & fid & c-acc & c-cov & e-fth & e-sc & e-con & lang \\
\midrule
Prism-Reranker-0.8B & .814 & .995 & .977 & 3.60 & 3.50 & 3.44 & 3.31 & 3.11 & 4.98 \\
Prism-Reranker-2B   & .828 & .996 & .968 & 3.75 & 3.65 & 3.63 & 3.43 & 3.17 & 4.93 \\
Prism-Reranker-4B   & .834 & .997 & .975 & 3.85 & 3.80 & 3.71 & 3.56 & 3.31 & 4.99 \\
Prism-Reranker-9B   & .845 & .999 & .972 & 3.88 & 3.82 & 3.70 & 3.57 & 3.33 & 4.95 \\
\midrule
Prism-Reranker-4B-exp & .851 & .996 & .975 & 3.75 & 3.66 & 3.55 & 3.35 & 3.29 & 4.94 \\
\midrule
\textsc{DeepSeek-V4-Flash}$^{\dagger}$ & n/a & n/a & .914 & 3.93 & 3.90 & 3.76 & 3.72 & 3.57 & 4.92 \\
\bottomrule
\end{tabular}
\end{table}

Table~\ref{tab:ce} summarizes the results across all nine dimensions.
Format compliance is near-perfect ($\ge 0.995$) and language consistency exceeds 4.93 for every model, indicating that the structured output protocol is reliably followed regardless of model size. Label accuracy scales steadily from .814 (0.8B) to .845 (9B), and all six LLM-judged quality dimensions follow the same trend, with the largest gains in contribution coverage ($+$0.32 from 0.8B to 9B) and evidence self-containedness ($+$0.26). Entity fidelity remains above .968 across the board, suggesting that hallucinated content is rare even for the smallest model.

Prism-Reranker-4B-exp achieves the highest label accuracy (.851) among all five models, which is expected given that its base model \textsc{Qwen3-Reranker-4B} was already fine-tuned for relevance classification. Its LLM-judged quality scores are slightly below those of Prism-Reranker-4B (which shares the same parameter count but was trained from scratch on the full distillation pipeline), most likely because the extension variant's SFT data uses self-generated teacher labels rather than the commercial teacher's scores. Nevertheless, all dimensions remain comfortably above the ``acceptable'' anchor of~3, confirming that the extension recipe produces usable contribution and evidence outputs.

\begin{figure}[t]
  \centering
  \includegraphics[width=\linewidth]{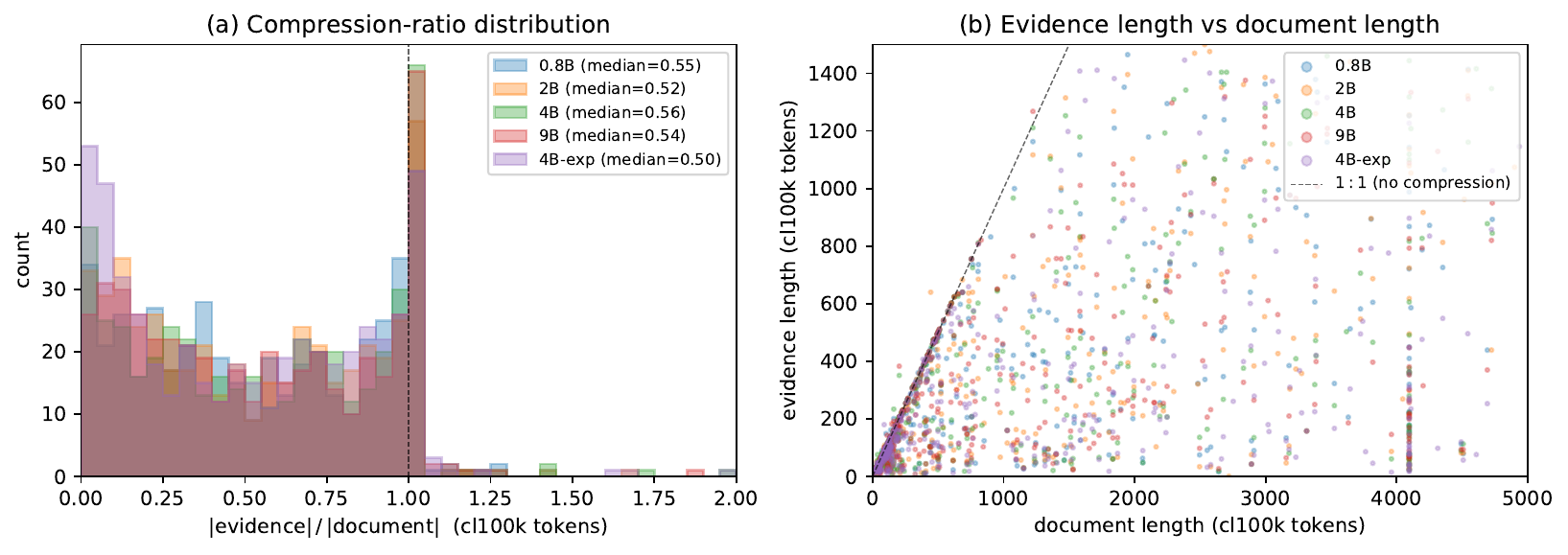}
  \caption{Compression statistics of the evidence field on the held-out dev set ($\sim$470 \texttt{yes}-labeled pairs per model, cl100k tokens). \textbf{(a)}~Per-pair compression-ratio distribution $|\mathrm{evidence}|/|\mathrm{document}|$, clipped at 2.0 for display; the dashed line marks $r=1$ (no compression). \textbf{(b)}~Per-pair scatter of evidence length against document length; the dashed diagonal marks the $1{:}1$ no-compression line.}
  \label{fig:compression}
\end{figure}

\paragraph{Compression ratio.} Beyond per-dimension quality, the practical value of the evidence field depends on how much it shortens the source document. \cref{fig:compression} reports the per-pair compression ratio $r = |\mathrm{evidence}| / |\mathrm{document}|$ measured in cl100k tokens over all dev-set pairs the model labels \texttt{yes}. The median ratio is approximately $0.5$ across all five released models---0.55 (0.8B), 0.53 (2B), 0.56 (4B), 0.54 (9B), 0.50 (4B-exp)---meaning the evidence field is on average half the length of the source document. The 10th-percentile ratio falls to $\sim 0.07$, which corresponds to long noisy web pages condensed to a single relevant span; the 90th percentile saturates near $1.0$, where short, already-concise documents are preserved nearly verbatim. The scatter in panel~(b) confirms that evidence length grows sub-linearly in document length, so the largest absolute token savings accrue precisely on the longest inputs---the regime where context-length pressure on downstream LLMs is greatest.

\paragraph{Strong-LLM baseline.} As an external reference point, we additionally evaluate \textsc{DeepSeek-V4-Flash}---a 284B-total / 13B-active-parameter mixture-of-experts model, roughly $32\times$ the total parameters of Prism-Reranker-9B---prompted with the same input and asked to emit contribution and evidence in the same structured format. We deliberately do \emph{not} use the larger sibling \textsc{V4-Pro} ($\sim$1.6T parameters) for this comparison: V4-Pro is itself the generator of our SFT contribution/evidence targets (\cref{sec:data:ce}), and benchmarking the student directly against its own training teacher would not be informative. V4-Flash, while sharing a model family, is a substantially smaller and weaker model than V4-Pro and serves as a deployable proxy ceiling rather than as the true upper bound; the true upper bound is V4-Pro itself, and any number a student reports on the deterministic entity-fidelity metric should be read against \emph{V4-Pro's} quality, not V4-Flash's.

On entity fidelity, computed by deterministic substring matching against the source document, Prism-Reranker-9B reports $.972$ versus V4-Flash's $.914$. We do not interpret this as the student outperforming its teacher: it merely reflects that V4-Flash---a much smaller model than V4-Pro---paraphrases the source more aggressively, and the gap to V4-Pro itself, which we have not measured directly, is plausibly far smaller or in the other direction. On five of the six LLM-judged dimensions Flash leads Prism-Reranker-9B by $0.05$--$0.24$ points on the 1--5 scale; on the sixth---language consistency---Prism-Reranker-9B holds a marginal lead ($4.95$ vs $4.92$). We openly acknowledge that on the subjective dimensions our compact models still trail a frontier-scale LLM. We position Prism-Reranker accordingly: not as a quality-at-any-cost competitor to frontier LLMs, but as a deployable family of dense $\le 9$B models that runs the full gate$\,+\,$contribution$\,+\,$evidence pipeline in a single forward pass on commodity GPUs, which is the regime that matters for low-budget and on-premise deployments where invoking a 284B-class (let alone 1.6T-class) API per query is impractical. Closing the remaining quality gap by further scaling both the backbone and the training corpus is left to future work. As flagged at the top of this section, the LLM-judged columns measure fidelity to the V4-Pro output style; \textsc{V4-Flash} shares a model family with the V4-Pro judge and may benefit from stylistic affinity, so its lead on those dimensions should not be over-interpreted.

\subsection{Distillation Loss Choice}
\label{sec:experiments:ablation}

Our method uses point-wise MSE as the sole distillation component (\cref{sec:method:train}). We justify this choice with a controlled ablation on a shared \textsc{Qwen3-Reranker-0.6B}~\citep{zhang2025qwen3emb} backbone, varying only the distillation loss across $\mathcal{L}_{\text{point}}$, $\mathcal{L}_{\text{list}}$, their combination, and a three-way combination with a weighted InfoNCE rank loss $\mathcal{L}_{\text{rank}}$~\citep{hofstatter2020marginmse,zhuang2023rankt5}. Evaluated on 80 retrieval datasets from MTEB~\citep{muennighoff2022mteb} and PosIR~\citep{zeng2026posir}, all four recipes improve over the non-distilled backbone by at least $+2.26$ NDCG@10, but adding $\mathcal{L}_{\text{list}}$ or $\mathcal{L}_{\text{rank}}$ on top of $\mathcal{L}_{\text{point}}$ yields no further gain and slightly dilutes the signal; point-wise alone attains the best overall mean. We attribute this to the cross-encoder's full query--document interaction at inference: the continuous teacher score already supplies a dense per-pair signal that dual-encoder-style contrastive objectives were designed to substitute for. We caveat that this ablation runs on \textsc{Qwen3-Reranker-0.6B} rather than the Qwen3.5 backbone of our main models; we adopt point-wise distillation for Prism-Reranker on the basis of this signal but leave a Qwen3.5-backbone replication to future work. Full setup, grouped results, and per-MTEB-dataset numbers are in \cref{app:loss-ablation}.

\section{Discussion}
\label{sec:discussion}

\subsection{Is Reasoning Necessary for Reranking?}
\label{sec:discussion:reasoning}

Recent work on reasoning-augmented language models has prompted the question of whether explicit chain-of-thought (CoT) reasoning can improve passage reranking.
Prism-Reranker deliberately adopts a \emph{non-reasoning} architecture: the relevance decision is made via a single-token yes/no prediction followed by direct generation of contribution and evidence, with no intermediate reasoning trace.
We now situate this choice within the emerging empirical evidence.

\citet{jedidi2025overthink} conduct a controlled comparison between a reasoning-based pointwise reranker (ReasonRR) and a standard non-reasoning counterpart (StandardRR) trained under identical conditions.
StandardRR consistently outperforms ReasonRR; more strikingly, disabling the reasoning trace at inference time (ReasonRR-NoReason) yields better scores than the full reasoning variant.
The authors attribute this to a \emph{polarization effect}: the reasoning process pushes relevance scores toward the extremes, undermining the model's ability to capture partial relevance---precisely the fine-grained signal that pointwise rerankers rely on.

\citet{lu2025rethinking} present the first systematic study covering both pointwise and listwise rerankers, direct-output and reasoning-augmented variants, and both SFT and RL training paradigms.
Across the BEIR benchmark and the reasoning-intensive BRIGHT benchmark, reasoning-augmented rerankers \emph{consistently underperform} their direct-prediction counterparts, with NDCG@10 gaps of up to 9.0 points on BRIGHT---despite the substantially higher inference cost of generating a full reasoning chain.

These findings align with a broader cognitive-science perspective.
\citet{liu2024cot} demonstrate that, analogous to tasks where deliberate reasoning harms human performance, CoT can degrade state-of-the-art models on certain task families, with accuracy drops of up to 36.3\% for o1-preview relative to GPT-4o.
Relevance assessment appears to be one such task: it depends heavily on soft, holistic matching rather than multi-step logical deduction, and injecting an explicit reasoning trace can override the model's implicit pattern-matching strengths.

For Prism-Reranker, the implication is two-fold.
First, the direct yes/no scoring mechanism preserves the continuous relevance signal that is critical for accurate ranking, avoiding the polarization artifact observed with CoT.
Second, by not generating a reasoning chain, the model reserves its generation budget entirely for the \texttt{<contribution>} and \texttt{<evidence>} fields---outputs that provide concrete downstream value for agentic pipelines---rather than spending tokens on an intermediate rationale that, as shown above, would likely \emph{hurt} rather than help.

\subsection{The Case for Reinforcement Learning}
\label{sec:discussion:rl}

Prism-Reranker is trained entirely with supervised objectives (distillation loss plus SFT). While our evaluation of contribution and evidence quality (\cref{sec:experiments:ce}) shows satisfactory results under the current recipe, we believe reinforcement learning (RL) represents a promising direction for further improvement.

\paragraph{Task-level reward signals are naturally available.}
Several quality dimensions of the structured output lend themselves to well-defined, automatable reward functions:
(i)~\emph{conciseness}---whether the generated text avoids redundant phrasing and stays within a target length;
(ii)~\emph{language consistency}---whether the output language matches the document or query language;
(iii)~\emph{entity fidelity}---whether named entities, numbers, and dates in the evidence are verbatim copies from the source document rather than paraphrased or fabricated.
These criteria can be evaluated with lightweight rule-based checkers, making them well-suited as reward signals for RL without requiring an expensive LLM judge in the training loop.

\paragraph{Hallucination suppression.}
The most compelling motivation for RL is the elimination of hallucinated content in the evidence field. Under SFT alone, the model learns to \emph{imitate} teacher outputs and may generalize by producing plausible-sounding but unfaithful details---particularly when the source document is long and the relevant span is small. An RL objective that directly penalizes entity-level hallucination could teach the model a stronger invariant: \emph{never fabricate content that is absent from the source}. This is especially important in agentic settings where the downstream model trusts the evidence as a faithful proxy for the original document and has no opportunity to verify against the source.

\paragraph{Why we did not pursue RL in this work.}
The entire Prism-Reranker project---data curation, training, evaluation, and paper writing---was carried out by a single independent researcher with limited computational resources. Implementing a stable RL pipeline (reward model design, PPO or GRPO infrastructure, hyperparameter search) constitutes a substantial engineering effort that was beyond the scope of this release. We leave RL-based refinement of the structured outputs as future work, and we expect it to yield measurable gains in evidence faithfulness and conciseness.

\subsection{Flexible Training Methodology}
\label{sec:discussion:flexible}

The default Prism-Reranker recipe distills a commercial teacher while simultaneously training structured outputs. This section discusses two alternative scenarios that broaden the applicability of the framework.

\paragraph{Scenario 1: Augmenting an existing LLM-based reranker.}
When a reranker is architecturally a causal LM---as is the case for \textsc{Qwen3-Reranker}~\citep{zhang2025qwen3emb}---contribution and evidence generation can be grafted on through the extension variant described in \cref{sec:method:train}, with no external teacher required. The self-distillation objective anchors the model's ranking behaviour to its own pre-training checkpoint, while SFT teaches the new structured outputs.

Prism-Reranker-4B-exp provides concrete evidence for this scenario. Starting from \textsc{Qwen3-Reranker-4B}, the extension training improves average BEIR-QA NDCG@10 by $+$1.54 (Table~\ref{tab:beir}) while equipping the model with contribution and evidence capabilities whose quality scores sit comfortably above the acceptable threshold (Table~\ref{tab:ce}). As discussed in \cref{sec:method:train,sec:experiments:beir}, the self-distillation term carries no ranking information beyond what the base model already encodes---its loss-minimizing solution is simply to reproduce the frozen anchor---so the ranking gain in this variant is most plausibly driven by the SFT branch's ensemble-label supervision, possibly with additional regularisation from the joint contribution/evidence targets. The key prerequisite is that the base reranker must be a generative LM---encoder-only cross-encoders such as \textsc{bge-reranker-v2-m3}~\citep{chen2024bgem3} cannot produce free-form text and are therefore not amenable to this approach.

\paragraph{Scenario 2: Training from scratch without a commercial teacher.}
When no commercial reranker API is available for distillation, a two-stage curriculum offers a plausible path. In the first stage, the model is trained on ranking data using standard point-wise or list-wise objectives, with a small fraction of SFT examples mixed in to prevent the model from losing its text-generation capability. \citet{zhang2025qwen3emb} report that their reranker required large-scale data to achieve competitive ranking quality, suggesting that ranking is a data-intensive skill that benefits from being learned first. In the second stage, once ranking ability has converged, the model undergoes the same self-distillation-plus-SFT extension as Scenario~1 to acquire contribution and evidence outputs.

We have not validated this two-stage recipe experimentally; the description above is a methodological extrapolation from our observations on Scenario~1 and from the data requirements documented by \citet{zhang2025qwen3emb}. We present it here as a discussion point rather than a verified result, and leave empirical validation to future work.

\subsection{Other Limitations and Future Work}
\label{sec:discussion:other}

Beyond the RL direction discussed in \cref{sec:discussion:rl}, two further limitations of this release deserve explicit mention.

\paragraph{Ablation coverage.} The only methodological ablation we report is the choice of distillation loss (\cref{sec:experiments:ablation}). Several other design decisions---the 5-judge ensemble (compared against a single judge or against the teacher alone), the length--score balancing of the training corpus, the $\sim$30\% rate of keyword-style query rewrites, and the relatively aggressive choice of $\gamma_{\text{point}}=20$---are presented without controlled comparisons. We leave systematic ablations of these factors to future work.

\paragraph{End-to-end agentic evaluation.} The paper is framed around agentic retrieval, but all reported experiments are intrinsic: BEIR NDCG@10 measures ranking quality, and the contribution-and-evidence evaluation measures output quality through an LLM judge. We do not yet measure downstream task success when a language-model agent consumes Prism-Reranker's evidence field in place of full retrieved documents (e.g.\ open-domain QA exact-match, or prompt-token reduction at fixed downstream accuracy). Such an end-to-end study is the most direct test of the value proposition advertised in this paper and is the highest-priority item we leave for future work.

\paragraph{Faithfulness via SFT alone.} Evidence faithfulness, our most safety-critical output dimension, is enforced solely by the supervised fine-tuning loss. We do not employ constrained decoding, copy-bias mechanisms, or post-hoc entity verification at inference time. Our entity-fidelity numbers (\cref{tab:ce}) and the LLM judge's \texttt{evidence\_faithfulness} scores indicate that hallucinated tokens are rare in practice, but autoregressive generation provides no architectural guarantee against them. Reinforcement-learning rewards on entity-level fidelity, discussed in \cref{sec:discussion:rl}, are the most promising mitigation we are aware of and are explicitly left for future work.

\paragraph{Language coverage and data licensing.} The training corpus is dominated by English and Chinese, with smaller proportions of other languages inherited from the KaLM-Embedding aggregation~\citep{kalm2025dataset}. We have not measured ranking or contribution/evidence quality on truly low-resource languages, and the released models should be assumed strongest in the two majority languages. Separately, the web-sourced documents collected via Tavily and Exa carry per-page licensing terms that we cannot enumerate exhaustively; consumers of the released training corpus should treat the web slice as research-only and consult the original source URLs before redistributing any individual document.

\section{Conclusion}
\label{sec:conclusion}

We introduced Prism-Reranker, a family of open cross-encoder rerankers that extend the standard relevance-scoring interface with two additional outputs: a \texttt{<contribution>} sentence summarizing how a document helps the query, and an \texttt{<evidence>} passage that is a self-contained, query-focused distillation of the document's relevant content. Both outputs are produced in a single forward pass at negligible cost beyond the relevance score itself.

The training methodology combines point-wise distillation from a strong commercial reranker with supervised fine-tuning on LLM-generated structured targets, applied jointly to every training sample under a single combined loss. The role of an independently-constructed five-model LLM-as-Judge ensemble is to convert the teacher's continuous score into a clean binary tag that is consistent across heterogeneous open corpora; positives receive a full \texttt{<contribution>}/\texttt{<evidence>} SFT target, negatives receive a single \texttt{no} token. This binary signal, together with length--score balancing of the training corpus, yields a recipe under which the model produces high-quality contribution and evidence outputs at the cost of a few NDCG@10 points relative to the commercial teacher on BEIR-QA. The same recipe applied to an existing strong open-source reranker (Prism-Reranker-4B-exp) instead lifts ranking quality while adding the structured-output capability, suggesting that the ranking gap on the four Qwen3.5-based sizes is a multi-task tradeoff against a single-task teacher rather than a methodological ceiling.

Beyond the four Qwen3.5-based models, we show that the same training recipe readily extends to existing LLM-based rerankers: Prism-Reranker-4B-exp augments \textsc{Qwen3-Reranker-4B} with contribution and evidence outputs while improving its average BEIR-QA NDCG@10 by $+1.54$ over the base model, demonstrating that the approach is not tied to a single backbone or training-from-scratch workflow.

We hope that pairing a relevance signal with a faithful, compact evidence passage lowers the barrier between retrieval and reasoning in agentic pipelines: downstream language models receive precisely the information that is relevant to the query, reducing both prompt length and the risk of hallucination from noisy retrieved content.

\bibliographystyle{plainnat}
\bibliography{prism-refs}

\appendix
\section{Prompt Template and Output Format}
\label{app:prompt}

This appendix gives the exact training-time prompt and a worked input/output example. Inference uses the same template.

\subsection{Raw Template}

We feed the backbone with a raw string template rather than calling \texttt{apply\_chat\_template()}, so that the prompt boundary used for relevance scoring (\cref{eq:score}) is byte-identical across samples. The template is:

\begin{quote}\small\ttfamily
<|im\_start|>system\\
\{system\_prompt\}<|im\_end|>\\
<|im\_start|>user\\
<Instruct>: \{instruction\}\\
<Query>: \{query\}\\
<Document>: \{doc\}<|im\_end|>\\
<|im\_start|>assistant\\
<think>\\
\\
</think>\\
\\
\end{quote}

The \texttt{system\_prompt} is the single sentence ``Judge whether the Document meets the requirements based on the Query and the Instruct provided.'' The \texttt{instruction} field is the same for every sample:

\begin{quote}\small
Given a query and a document, judge whether the document is relevant to the query. Answer ``yes'' or ``no'', then provide in XML:
\begin{enumerate}
  \item \texttt{<contribution>}: what the document contributes to the query.
  \item \texttt{<evidence>}: a self-contained rewrite of relevant content.
\end{enumerate}
\end{quote}

The empty \texttt{<think></think>} block is intentional: it disables the backbone's chain-of-thought channel at training time, so the very next decoded token is the verdict. This makes the position of $\ell_{\texttt{yes}}$ and $\ell_{\texttt{no}}$ in \cref{eq:score} deterministic.

\subsection{Worked Examples}

We give two positive examples in different languages, then describe the negative case. Training data is multilingual, with Chinese and English as the dominant languages.

\paragraph{English example --- input.}
\begin{quote}\small
\textbf{Query:} How can I lose weight quickly?\\
\textbf{Document:} A 12-week randomized controlled trial ($n{=}200$) found that the intermittent fasting group lost on average $6.8$\,kg, significantly more than the $4.1$\,kg lost by the traditional caloric restriction group ($p<0.01$). The authors attribute the gap to a longer fat-oxidation window during the fasting periods.
\end{quote}

\paragraph{English example --- target output.}
\begin{quote}\small\ttfamily
yes\\
<contribution>Provides controlled-trial evidence that intermittent fasting outperforms traditional caloric restriction for short-term weight loss.</contribution>\\
<evidence>A 12-week randomized controlled trial with 200 participants found the intermittent fasting group lost 6.8 kg on average versus 4.1 kg for traditional caloric restriction (p<0.01); the authors attribute the gap to a longer fat-oxidation window.</evidence>
\end{quote}

\paragraph{Chinese example --- input.}
\begin{quote}\small
\textbf{Query:} Transformer 模型中的位置编码是干什么用的\\
\textbf{Document:} Transformer 是 Google 在 2017 年提出的神经网络架构，已被广泛应用于各类自然语言处理任务。由于自注意力机制本身不区分输入序列中元素的顺序，模型无法感知 ``我吃苹果'' 和 ``苹果吃我'' 的差异，因此需要引入位置编码（positional encoding）将位置信息注入输入向量。原始论文采用正弦余弦函数生成固定的位置编码，后续工作如 RoPE、ALiBi 等则提出了可学习或相对位置的方案。
\end{quote}

\paragraph{Chinese example --- target output.}
\begin{quote}\small\ttfamily
yes\\
<contribution>说明了位置编码的作用是为自注意力机制注入序列顺序信息，并列举了几种主流实现方式。</contribution>\\
<evidence>Transformer 的自注意力机制不区分输入元素顺序，需要位置编码将位置信息注入输入向量。原始论文使用正弦余弦函数生成固定位置编码；后续工作如 RoPE、ALiBi 提出了可学习或相对位置的方案。</evidence>
\end{quote}

In both cases the \texttt{<evidence>} field drops opening boilerplate (the trial's design framing in English, the historical attribution to Google in Chinese) while preserving every fact that bears on the query.

\paragraph{Negative example.}
When the document is irrelevant to the query, the target is the single token \texttt{no} with no contribution or evidence emitted.

\section{Distillation Loss Ablation}
\label{app:loss-ablation}

This appendix expands \cref{sec:experiments:ablation}: a controlled study of which distillation loss---or which combination of losses---best suits a generative cross-encoder student. Prior work has used list-wise KL against the teacher's rank distribution~\citep{zhuang2023rankt5}, pair-wise or InfoNCE-style rank losses~\citep{hofstatter2020marginmse}, and point-wise score regression; it is not a priori clear which is best suited here.

\paragraph{Setup.}
To separate the loss question from backbone choice, this ablation uses \textsc{Qwen3-Reranker-0.6B}~\citep{zhang2025qwen3emb} as \emph{both} the student and the no-distillation reference. All four distilled variants share the same backbone, the same commercial teacher (\cref{sec:method:train}), the same training data, and the same optimizer schedule; only the distillation loss changes. We compare four recipes:
\begin{itemize}
  \item $\mathcal{L}_{\text{point}}$: the point-wise MSE objective of \cref{eq:loss-point}.
  \item $\mathcal{L}_{\text{list}}$: Hinton-style KL divergence between the student and teacher score distributions over the 8-document training group (one positive, seven negatives). Teacher scores are mapped back to logits via inverse sigmoid, both sides are softened with temperature $T{=}2$, and the loss is multiplied by $T^{2}$ to restore gradient scale.
  \item $\mathcal{L}_{\text{point}} + \mathcal{L}_{\text{list}}$: the two combined with equal weight.
  \item $\mathcal{L}_{\text{point}} + \mathcal{L}_{\text{list}} + \mathcal{L}_{\text{rank}}$: additionally adds a weighted InfoNCE rank loss. Each negative is reweighted by its teacher-margin gap to the positive, so hard negatives (small margin) receive more gradient than trivial ones.
\end{itemize}
No SFT branch is used in any configuration, so this ablation speaks only to the choice of distillation loss and is orthogonal to the point+SFT recipe of our main model.

\paragraph{Evaluation.}
We evaluate on three heterogeneous benchmark groups: (\textit{i})~18 retrieval datasets from MTEB~\citep{muennighoff2022mteb} covering English and Chinese; (\textit{ii})~the Chinese subset of PosIR~\citep{zeng2026posir} (31 domain-stratified datasets spanning aerospace, biomedicine, finance, law, and 27 other domains); and (\textit{iii})~the English subset of PosIR (31 domains in parallel to the Chinese subset). For every (query, corpus) pair we take the top-100 candidates returned by the lightweight multilingual retriever \texttt{static-similarity-mrl-multilingual-v1}~\citep{staticmrl}; if the annotated positive is not among them we force-insert it, so that the metric reflects reranking quality alone and is not upper-bounded by first-stage recall. We report NDCG@10.

\begin{table}[t]
\centering
\caption{NDCG@10 (\%) averaged within each benchmark group. All four distilled variants share the same \textsc{Qwen3-Reranker-0.6B} backbone, teacher, and training data; only the distillation loss differs. Best per column in \textbf{bold}. MTEB: 18 datasets; PosIR-zh / PosIR-en: 31 Chinese / English domain-stratified datasets each; Overall: unweighted mean over all 80 datasets.}
\label{tab:ablation-loss}
\small
\setlength{\tabcolsep}{8pt}
\begin{tabular}{lcccc}
\toprule
Configuration & MTEB & PosIR-zh & PosIR-en & Overall \\
\midrule
\textsc{Qwen3-Reranker-0.6B} (no distillation) & 82.11 & 94.73 & 95.23 & 92.08 \\
\midrule
$+\ \mathcal{L}_{\text{list}}$                                                              & 84.98          & 97.24          & 96.88          & 94.34 \\
$+\ \mathcal{L}_{\text{point}}$                                                             & \textbf{85.08} & \textbf{97.58} & 97.04          & \textbf{94.56} \\
$+\ \mathcal{L}_{\text{point}} + \mathcal{L}_{\text{list}}$                                 & \textbf{85.08} & 97.44          & 96.97          & 94.48 \\
$+\ \mathcal{L}_{\text{point}} + \mathcal{L}_{\text{list}} + \mathcal{L}_{\text{rank}}$     & 85.02          & 97.42          & \textbf{97.07} & 94.49 \\
\bottomrule
\end{tabular}
\end{table}

\begin{table}[t]
\centering
\caption{Per-dataset NDCG@10 (\%) on the 18 MTEB retrieval datasets. Same settings as \cref{tab:ablation-loss}. \textbf{L}, \textbf{P}, \textbf{R} denote $\mathcal{L}_{\text{list}}$, $\mathcal{L}_{\text{point}}$, $\mathcal{L}_{\text{rank}}$; the first column is the non-distilled backbone.}
\label{tab:ablation-mteb}
\small
\setlength{\tabcolsep}{5pt}
\begin{tabular}{lccccc}
\toprule
Dataset & Qwen3-RR-0.6B & $+\,$L & $+\,$P & $+\,$P$+$L & $+\,$P$+$L$+$R \\
\midrule
ClimateFEVER        & 63.15 & 68.87 & 69.43 & 69.93 & 69.46 \\
CmedqaRetrieval     & 68.38 & 74.48 & 76.31 & 75.61 & 75.21 \\
CovidRetrieval      & 95.54 & 98.02 & 98.02 & 98.32 & 98.39 \\
DuRetrieval         & 96.17 & 96.47 & 95.91 & 96.49 & 96.01 \\
EcomRetrieval       & 82.39 & 84.74 & 85.28 & 84.85 & 85.78 \\
FEVER               & 95.01 & 96.36 & 96.81 & 96.67 & 96.63 \\
HotpotQA            & 90.15 & 92.01 & 92.49 & 91.91 & 92.27 \\
MMarcoRetrieval     & 89.86 & 89.09 & 89.61 & 89.34 & 88.81 \\
MedicalRetrieval    & 80.00 & 89.18 & 87.91 & 89.67 & 87.82 \\
T2Retrieval         & 94.66 & 97.80 & 97.87 & 97.77 & 97.89 \\
VideoRetrieval      & 85.49 & 89.07 & 89.10 & 87.78 & 89.14 \\
ArguAna             & 77.82 & 81.31 & 81.70 & 82.10 & 82.27 \\
CQADupStack Gaming  & 82.02 & 82.34 & 81.98 & 81.91 & 82.09 \\
CQADupStack Unix    & 74.07 & 77.03 & 77.34 & 77.48 & 77.32 \\
FiQA                & 69.96 & 69.26 & 68.97 & 67.99 & 67.79 \\
SCIDOCS             & 44.62 & 51.80 & 52.15 & 51.83 & 51.85 \\
TREC-COVID          & 99.73 & 99.86 & 99.81 & 99.85 & 99.81 \\
Touch\'e 2020 v3    & 88.96 & 91.93 & 90.82 & 91.99 & 91.80 \\
\midrule
Average             & 82.11 & 84.98 & 85.08 & 85.08 & 85.02 \\
\bottomrule
\end{tabular}
\end{table}

\paragraph{Findings.}
\cref{tab:ablation-loss} shows that all four distillation recipes lift the backbone by a large and consistent margin---between $+2.26$ and $+2.48$ NDCG@10 overall---so distillation from the commercial teacher is useful regardless of loss shape. Among the four recipes the gaps are small (within $0.22$ point overall), but the simplest objective wins: $\mathcal{L}_{\text{point}}$ alone attains the highest overall mean and is best or tied on MTEB and PosIR-zh. Adding $\mathcal{L}_{\text{list}}$, or further adding $\mathcal{L}_{\text{rank}}$, does not help and slightly dilutes the signal. The per-dataset view in \cref{tab:ablation-mteb} is consistent: on 16 of 18 MTEB datasets point-wise distillation lands within $0.4$ NDCG@10 of the best recipe, and the only two datasets where the non-distilled backbone remains competitive (FiQA, MMarcoRetrieval) penalize all four recipes roughly equally---implicating a teacher--domain mismatch rather than a loss choice.

We attribute the sufficiency of point-wise distillation to the cross-encoder's full query--document interaction at inference time. Contrastive and list-wise losses were originally motivated by dual-encoder students whose bottleneck is representation capacity; for a cross-encoder, the continuous teacher score already supplies a dense per-pair supervision signal rich enough to recover the teacher's ranking without an auxiliary contrastive objective.

Per-dataset NDCG@10 figures for the 31 PosIR-zh and 31 PosIR-en domains underlying \cref{tab:ablation-loss} are not reproduced here for space; readers interested in the domain composition and per-domain protocol of the benchmark are referred to the original PosIR paper~\citep{zeng2026posir}.

\section{Qualitative Examples: Strengths and Failure Modes}
\label{app:cases}

This appendix presents four qualitative examples drawn from the held-out dev set, all generated by Prism-Reranker-9B and scored under the LLM-as-Judge pipeline of \cref{sec:experiments:ce}. The first two illustrate strengths of the structured output; the last two illustrate failure modes that future iterations should target.

Each example reports the query, the source document (truncated where necessary), the model's contribution and evidence outputs, and a short commentary.

\subsection{Strength: Extracting a Clean Answer from a Noisy Source (Chinese)}

\begin{description}
  \item[Query.] 燕子乌鸦老鹰小鸟麻雀哪个是不同类的词语
  \item[Document ($\sim$914 cl100k tokens, abbreviated).] 来源:学生作业帮助网 \dots\ 燕子乌鸦老鹰小鸟麻雀那个不是同一类 \dots\ 小鸟，因为小鸟是人对小型鸟类或鸟类幼畜的泛称，而剩下的都是不同鸟类的名字 \dots\ [followed by $\sim$750 characters of unrelated forum questions copied from the same page: 黑熊松鼠老鹰小鹿紫貂哪个不同类, 信鸽老鹰松鼠喜鹊海鸥, 蝈蝈蜥蜴蚂蚁甲虫瓢虫, etc.]
  \item[Contribution.] 指出``小鸟''是泛称而非具体鸟类名称，其余均为具体鸟种，因此``小鸟''是不同类的词语。
  \item[Evidence.] 小鸟是人对小型鸟类或鸟类幼畜的泛称，而燕子、乌鸦、老鹰、麻雀都是不同鸟类的名字，所以小鸟是不同类的词语。
\end{description}
\noindent The source is a forum scrape dominated by repeated copies of the question stem and unrelated similar questions; the model isolates the single answer-bearing span and rewrites it concisely. Judge scores: contribution\_accuracy~5, contribution\_coverage~5, evidence\_faithfulness~4.

\subsection{Strength: Concise Fact Extraction (English)}

\begin{description}
  \item[Query.] in what year was new york city divided into its boroughs?
  \item[Document.] title: 125 years ago, the five boroughs are joined to create today's New York City \textbar\ Welcome2TheBronx\\
  content: January 2, 2023 - On January 1st, 1898, after years of planning and razor-thin election results, the five boroughs we know today as The Bronx, Brooklyn, Manhattan, Queens, and Staten Island, were consolidated into one city to create the New York City we know
  \item[Contribution.] Provides the exact year when the five boroughs of New York City were consolidated into a single city.
  \item[Evidence.] On January 1st, 1898, the five boroughs---The Bronx, Brooklyn, Manhattan, Queens, and Staten Island---were consolidated into one city to create the New York City we know today.
\end{description}
\noindent The evidence drops the page-publication date ``January 2, 2023'' (irrelevant to the historical year the query asks about) while preserving the load-bearing fact ``January 1st, 1898''. All judge dimensions score~$\ge 4$.

\subsection{Failure Mode: Evidence Hallucination from a Short Source (Chinese)}

\begin{description}
  \item[Query.] 牡丹和芍药花的区别在哪里
  \item[Document ($\sim$125 cl100k tokens; truncated by the source webpage at ``多'').] title: 牡丹、芍药傻傻分不清楚？药师教您轻松辨别（内附食疗养生方 \dots\\
  content: 第一，来源相同。二者同为毛茛科植物，所以在外观上长得很相似，让人分不清楚。第二，牡丹花的层数多而厚，看起来雍容华贵的感觉，而芍药的花的层数比较少，多
  \item[Contribution.] 从植物来源、花朵层数与外观形态、花期时间、药用部位四个方面对比了牡丹和芍药的区别，并提供了食疗养生方。
  \item[Evidence.] 牡丹和芍药同为毛茛科植物，来源相同，外观相似。牡丹花的层数多而厚，看起来雍容华贵；芍药的花层数比较少。\textbf{牡丹花期在春季，芍药花期在夏季。牡丹药用部位是根皮，芍药药用部位是根。食疗养生方包括：牡丹花泡茶、芍药花泡茶、牡丹花煮粥、芍药花煮粥。}
\end{description}
\noindent The bolded portions of the evidence are not in the source. The document only mentions ``食疗养生方'' (dietary recipes) and ``花期'' (flowering period) inside its truncated title; it never specifies a season, a medicinal part, or any recipe. The model fills these gaps with plausible-sounding but unsupported content---a textbook case of the SFT-level hallucination discussed in \cref{sec:discussion:other}, which an entity-fidelity reward in an RL stage (\cref{sec:discussion:rl}) would directly penalize. Judge scores: evidence\_faithfulness~1, contribution\_accuracy~2.

\subsection{Failure Mode: Contribution Overreach (English)}

\begin{description}
  \item[Query.] Which element is added to iron to make steel?
  \item[Document.] It mostly contains iron, but chromium, carbon, nickel, and other elements are added to make it aesthetic, strong, durable, and corrosion resistant. We can
  \item[Contribution.] Identifies that chromium, carbon, nickel, and other elements are added to iron to make \emph{steel} with specific properties.
  \item[Evidence.] It mostly contains iron, but chromium, carbon, nickel, and other elements are added to make it aesthetic, strong, durable, and corrosion resistant.
\end{description}
\noindent The evidence is verbatim faithful to the document (entity-fidelity score $1.0$, evidence\_faithfulness~4). The contribution, however, asserts that the document is about \emph{steel}---a word that never appears in the snippet, which only describes an iron-based alloy by its properties. The model has implicitly answered the user's query rather than describing what the source contributes, conflating the two roles of the contribution field. Judge: contribution\_accuracy~2.

\end{document}